\documentclass{procmult}
\begin{document}
\title{Deformations of spacetime metrics}
\author{D. Pugliese$^{a}$$^{b}$, C. Stornaiolo$^{c}$ \and S. Capozziello$^{d}$ }
\institute{ $^{b}$  International Center for Relativistic
Astrophysics - I.C.R.A.,
  University of Rome ``La Sapienza'', I-00185 Rome, Italy \\
$^{a}$Physics Department, University of Rome ``La Sapienza,''
I-00185 Rome, Italy  \\
$^{d}$Dipartimento di Scienze Fisiche, Università di Napoli
``Federico II'', and \\
$^{c}$INFN, Sez. di Napoli, Compl. Univ. di Monte S. Angelo,
Edificio G, Via Cinthia, I-80126 - Napoli, Italy } \maketitle
 \abstract{This work is devoted to  study the deformation of
 spacetime
metrics as generalized  conformal transformations. Some applications
are also considered, in particular the equations of motion in
deformed spacetime are studied.}
\section{Introduction}
Spacetime deformations are interesting to study in geometrical as
well as in physical situations, as exact solutions of Einstein's
equations can be modified into new metrics which better describe the
physical observations. On the other hand since, according to the
Gauss law, given a two-dimensional metric $g$ it is always possible
find a scalar  field such that $g$ is locally conformally flat,
observing that conformal transformations are a particular case of
deformations, we show how Gauss law can be generalized  to every
couple of n-dimensional metrics $(g,h)$, deforming  one metric into
other by matrices of scalar fields. Finally, geodetic motions of
test particles into deformed manifold are considered, in particular
lensing effects and gravitational redshift are studied. The equation
of gravitational waves is also obtained in the case of ``small"
deformation of the background.
\section{Deformations}
In order to define the deformation of a metric $g$ on an
n-dimensional manifold $M$ \cite{Storny,Tesi}, let us decompose  it
with a vielbein $\omega_{a}^{\mbox{\tiny{C}}}\,
(e^{a}_{\mbox{\tiny{C}}}$) as
$g_{ab}=\eta_{\mbox{\tiny{AB}}}\omega_{a}^{\mbox{\tiny{A}}}\omega_{b}^{\mbox{\tiny{B}}}
$ ($
g^{ab}=\eta^{\mbox{\tiny{AB}}}e^{a}_{\mbox{\tiny{A}}}e^{b}_{\mbox{\tiny{B}}}
$). If $\phi^{ \mbox{\tiny{A}}}_{ \phantom\ \mbox{\tiny{B}}}$ is a
scalar fields matrix on $M$, we can define a deformed vielbein
$\tilde{\omega}^{\mbox{\tiny{B}}}_{a}\equiv\omega^{\mbox{\tiny{A}}}_{a}\phi^{\mbox{\tiny{B}}}_{\phantom\
\mbox{\tiny{A}}}$ and $
\tilde{e}_{\mbox{\tiny{A}}}^{a}\equiv\psi^{\phantom\
\mbox{\tiny{B}}}_{ \mbox{\tiny{A}}}e_{\mbox{\tiny{B}}}^{a}$ where
 $\psi^{\mbox{\tiny{A}}}_{\phantom\ \mbox{\tiny{B}}}=\phi^{-1
\mbox{\tiny{A}}}_{ \phantom\ \mbox{\tiny{B}}}$ (first deforming
matrix). The following commutation relations for the deformed basis
hold
$$
\left[\widetilde{e}_{\mbox{\tiny{A}}},\widetilde{e}_{\mbox{\tiny{B}}}\right]^{b}=\psi_{
\mbox{\tiny{A}}}^{\phantom\ \mbox{\tiny{S}}}\psi_{
\mbox{\tiny{B}}}^{\phantom\ \mbox{\tiny{T}}
}\left[e_{\mbox{\tiny{S}}},e_{\mbox{\tiny{T}}}\right]^{b}+2\psi_{[\mbox{\tiny{A}}}^{\phantom\
\mbox{\tiny{S}}}e_{|\mbox{\tiny{S}}}^{a}
e_{\mbox{\tiny{T}}}^{b}\left(\partial_{a|}\psi_{
\mbox{\tiny{B}}]}^{\phantom\ \mbox{\tiny{T}}}\right)\, $$
 A new
metric $\widetilde{g}$ is obtained  as follows
$$
\widetilde{g}_{ab}=\eta_{\mbox{\tiny{AB}}}
\widetilde{\omega}_{a}^{\mbox{\tiny{C}}}\widetilde{\omega}_{b}^{\mbox{\tiny{D}}},
\;\widetilde{g}^{ab}=\eta^{\mbox{\tiny{AB}}}
\tilde{e}^{a}_{\mbox{\tiny{C}}}\tilde{e}^{b}_{\mbox{\tiny{D}}}
\quad\mbox{or equivalently}\quad
\widetilde{g}_{ab}=\mathcal{G}_{\mbox{\tiny{CD}}}\omega_{a}^{\mbox{\tiny{C}}}\omega_{b}^{\mbox{\tiny{D}}},\quad
\widetilde{g}^{ab}=\mathcal{G}^{\mbox{\tiny{CD}}}e^{a}_{\mbox{\tiny{C}}}e^{b}_{\mbox{\tiny{D}}},
$$
where
$\mathcal{G}_{\mbox{\tiny{CD}}}\equiv\eta_{\mbox{\tiny{AB}}}\phi_{\phantom\
\mbox{\tiny{C}}}^{ \mbox{\tiny{A}}}\phi_{\phantom\
\mbox{\tiny{D}}}^{ \mbox{\tiny{B}}}$ and $
\mathcal{G}^{\mbox{\tiny{CD}}}\equiv\eta_{\mbox{\tiny{AB}}}\psi_{\mbox{\tiny{A}}}^{\phantom\
\mbox{\tiny{C}}}\psi^{\phantom\ \mbox{\tiny{D}}}_{\mbox{\tiny{B}}} $
(second deforming matrix). The set of first deforming matrices form
a right coset for right composition with Lorentz matrix in vielbeins
indices; the identity is the class of equivalence composed by the
Lorentz matrices. Deformations should not be confused with
diffeomorphisms. In order to understand better this point let us
write  the metric $
\widetilde{g}_{\alpha\beta}=g_{\mu\nu}\phi^{\mu}_{\phantom\
\alpha}\phi^{\nu}_{\phantom\ \beta}$ where $ \phi^{a}_{\phantom\
b}\equiv \phi^{\mbox{\tiny{A}}}_{\phantom\
 \mbox{\tiny{B}}}\omega^{\mbox{\tiny{B}}}_{b}e^{a}_{\mbox{\tiny{A}}}$,
 then
when the matrices  $\phi^{\mbox{\tiny{A}}}_{\phantom\
\mbox{\tiny{B}}}$  define  $\phi^{\alpha}_{\phantom\ \beta}$
jacobian  matrices then the metric $\widetilde{g}_{ab}$ and the
tensor $g_{ab}$ are equivalent .
\subsection{Deformations of three-dimensional metrics}
 Coll et al. showed in \cite{3--dim} that any
three-dimensional metric can be locally obtained as a conformal
transformation of a constant curvature metric $h$ summed to  the
tensor product of a 1-form by itself, or $g=\sigma h+\epsilon\;
\textbf{s}\otimes\textbf{s}$, where $\epsilon=\pm1$ and $\sigma$ is
a scalar . Moreover, according to Riemann's theorem,  a scalar
relation $\Psi(\sigma,\|\mathbf{s}\|)=0$ between $\sigma$ and
$\mathbf{s}$ has to be imposed  as  the metric can be defined, at
most, by three independent functions. We find the first deforming
matrix associated to the deformation  $g=\sigma h+\epsilon\;
\textbf{s}\otimes\textbf{s}$, as
\begin{equation}\label{fatecore}
\phi_{\phantom\
\mbox{\tiny{C}}}^{\mbox{\tiny{A}}}=\sqrt{\sigma}\delta_{\phantom\
\mbox{\tiny{C}}}^{\mbox{\tiny{A}}}+\alpha
s^{\mbox{\tiny{A}}}s_{\mbox{\tiny{C}}},\quad \mbox{where $\alpha$ is
an arbitrary function of $\textbf{s}$ and $\|\mathbf{s}\|$}.
\end{equation}
\subsection{Proprieties of deformed metrics}
Conformal deformations are a particular simple example of
deformations where  the first deforming matrix is $
\psi^{\mbox{\tiny{C}}}_{\phantom\
\mbox{\tiny{A}}}=\Omega\Lambda^{\mbox{\tiny{C}}}_{\phantom\
\mbox{\tiny{A}}}. $  More generally a metric $h$ can be deformed in
$g$ by
\begin{equation}\label{siamquapertuadifesa}
\widetilde{g}_{ab}= g_{ab} + h_{ab}, \quad \widetilde{g}^{ab}=
g^{ab} + X^{ab},
\end{equation}
where $ X^{ab}$ is a tensor. In terms of the second deforming
matrices $\mathcal{G}_{\mbox{\tiny{AB}}}$ and
$\mathcal{G^{\mbox{\tiny{AB}}}}$, the (\ref{siamquapertuadifesa})
reads
\begin{equation}\label{infelice}
\mathcal{G}_{\mbox{\tiny{AB}}}=\eta_{\mbox{\tiny{AB}}}
+\mathcal{C}_{\mbox{\tiny{AB}}},\quad
\mathcal{G}^{\mbox{\tiny{AB}}}=\eta^{\mbox{\tiny{AB}}}+\mathcal{D}^{\mbox{\tiny{AB}}}\quad\mbox{with}
\quad X^{ab}=\mathcal{D}^{\mbox{\tiny{AB}}}e_{\mbox{\tiny{A}}}^{a}
e^{b}_{\mbox{\tiny{B}}}.
\end{equation}
If $ X^{ad}=-g^{ab}\widetilde{g}^{cd}h_{bc} $
$(\mathcal{D}^{\mbox{\tiny{AD}}}=-\eta^{\mbox{\tiny{AB}}}\mathcal{G}^{\mbox{\tiny{CD}}}\mathcal{C}_{\mbox{\tiny{BC}}}
)$ then $X^{ab}h_{bc}=\delta^{a}_{c}
 $(
$\mathcal{D}^{\mbox{\tiny{AB}}}\mathcal{C}_{\mbox{\tiny{BC}}}=\delta^{\mbox{\tiny{A}}}_{\phantom\
\mbox{\tiny{C}}}$), otherwise  $ X^{ab}h_{bc}\neq\delta^{a}_{c}$. In
particular, given the matrix $\varphi$ defined by $
\phi^{\mbox{\tiny{A}}}_{\phantom\
\mbox{\tiny{B}}}=\delta^{\mbox{\tiny{A}}}_{\phantom\
\mbox{\tiny{B}}}+\varphi^{\mbox{\tiny{A}}}_{\phantom\
\mbox{\tiny{B}}}$, the following relation holds
$$
\mathcal{G}_{\mbox{\tiny{AB}}}=\eta_{\mbox{\tiny{CD}}}\phi^{\mbox{\tiny{C}}}_{\phantom\
\mbox{\tiny{A}}}\phi^{\mbox{\tiny{D}}}_{\phantom\
\mbox{\tiny{B}}}=\eta_{\mbox{\tiny{AB}}}+C_{\mbox{\tiny{AB}}}
\quad\mbox{ with}\quad
C_{\mbox{\tiny{AB}}}=\eta_{\mbox{\tiny{CD}}}\varphi^{\mbox{\tiny{C}}}_{\phantom\
\mbox{\tiny{A}}}\varphi^{\mbox{\tiny{D}}}_{\phantom\
\mbox{\tiny{B}}}$$, if $ \varphi_{\phantom\
\mbox{\tiny{A}}}^{\mbox{\tiny{A}}}=0$, $ \varphi_{\phantom\
\mbox{\tiny{0}}}^{\mbox{\tiny{A}}}=\varphi_{\phantom\
\mbox{\tiny{A}}}^{\mbox{\tiny{0}}}$, $\varphi_{\phantom\
\mbox{\tiny{i}}}^{\mbox{\tiny{A}}}=-\varphi_{\phantom\
\mbox{\tiny{A}}}^{\mbox{\tiny{i}}}$ where $ i\in\left\{1,2,3\right\}
$ or
 $ \varphi_{\phantom\
\mbox{\tiny{A}}}^{\mbox{\tiny{B}}}=\left(
\begin{array}{cccc}
0 & a& b &c  \\
 a & 0&d_{1}&d_{2}\\
 b & -d_{1}&0&d_{3}\\
c&-d_{2}&-d_{3}&0\\
 \end {array}
 \right)
$ where  $a,  b, c, d_{1},  d_{2}, d_{3}$ are six scalar fields
defined on the manifold $M$, these are  just the right number of
degrees of freedom for a Riemannian 4--dimensional metric. More
generally, deformation
\begin{equation}\label{lassassinoiltrucido}
\widetilde{g}_{ab}=\Omega^{2} g_{ab}+ h_{ab},\quad
\widetilde{g}^{ab}=\Omega^{-2}g^{ab}+X^{ab}\quad\mbox{where}\quad
X^{ad}\equiv-\Omega^{-2}g^{ab} \widetilde{g}^{cd} h_{bc}
\end{equation}
$ h_{ab}\equiv\eta_{\mbox{\tiny{AB}}}\varphi_{\phantom\
\mbox{\tiny{C}}}^{\mbox{\tiny{A}}}\varphi_{\phantom\
\mbox{\tiny{D}}}
 ^{\phantom\ \mbox{\tiny{B}}}\
\omega^{\mbox{\tiny{C}}}_{a}\omega^{\mbox{\tiny{D}}}_{b}$  with $
\phi^{\mbox{\tiny{A}}}_{\phantom\ \mbox{\tiny{B}}}=\Omega
\delta^{\mbox{\tiny{A}}}_{\phantom\
\mbox{\tiny{B}}}+\varphi^{\mbox{\tiny{A}}}_{\phantom\
\mbox{\tiny{B}}}$  extends  the  (\ref{fatecore}), and the Gauss law
to any couple of four--dimension metrics $(g,h)$.
\subsection{Gravitational waves}
We consider the metric (\ref{siamquapertuadifesa}) as perturbation
of the exact metric $g$, solution of $G_{ab}=0$,  if the scalar
fields satisfy  the following condition $ \hbox{ $ | \
\mathcal{C}_{\mbox{\tiny{AB}}}|\ll1\  \ \forall
\mbox{\footnotesize{A}}\ e\ \mbox{\footnotesize{B}}$} , $ where $| \
\mathcal{C}_{\mbox{\tiny{AB}}}|=|\eta_{\mbox{\tiny{CD}}}\varphi^{\mbox{\tiny{C}}}_{\phantom
\ \mbox{\tiny{A}}}\varphi^{\mbox{\tiny{D}}}_{\phantom \
\mbox{\tiny{B}}}|$\ is the module of the matrix element
$\mathcal{C}_{\mbox{\tiny{AB}}}$(see also \cite{Storny}). Therefore
the perturbation of the metric $g$ is given by the tensor $h_{ab}$
by the following: $ \widetilde{g}_{ab}=g_{ab}+h_{ab}, $ while, since
in the given approximation we have neglected the terms
$X^{ad}h_{dc}$ and $h_{ab}h^{bd}$, we have $
\widetilde{g}^{ab}=g^{ab}-h^{ab}$. After some manipulation, adopting
the same gauge choice as in \cite{Wald} we obtain the wave equation
for the perturbation $h_{ac}$ in the form:
\begin{equation}\label{Rischi}
\nabla^{b} \nabla_{b}  h_{ac}=2R^{b\phantom\ \phantom\ d}_{\phantom\
ac}h_{bd}, \quad \nabla_{b}  g_{ac}=0.
\end{equation}
From (\ref{Rischi}) we finally find the following wave equation:
\begin{equation}\label{cor}
\left(\nabla^{b} \nabla_{b}
\mathcal{C}_{\mbox{\tiny{AB}}}\right)\omega^{\mbox{\tiny{A}}}_{\phantom\
a}\omega^{\mbox{\tiny{B}}}_{\phantom\ c}=
\mathcal{C}_{\mbox{\tiny{AB}}}\left[2R^{b\phantom\ \phantom\
d}_{\phantom\ ac}\omega^{\mbox{\tiny{A}}}_{\phantom\
b}\omega^{\mbox{\tiny{B}}}_{\phantom\ d}-\nabla^{b}
\nabla_{b}\left(\omega^{\mbox{\tiny{A}}}_{\phantom\
a}\omega^{\mbox{\tiny{B}}}_{\phantom\ c}\right)\right]
\end{equation}
In particular, for  deformation of the flat manifold the (\ref{cor})
is
$(\nabla^{b}\nabla_{b}\mathcal{C}_{\mbox{\tiny{AB}}})\omega^{\mbox{\tiny{A}}}_{\phantom\
a}\omega^{\mbox{\tiny{B}}}_{\phantom\ b}=0$.
\subsection{Geodetic motion}
In order to consider the geodetics in  deformed spacetime let us
take into account the metric (\ref{siamquapertuadifesa}). The
geodetic equation in deformed spacetime is
\begin{equation}\label{amiciaddio}
v^{a}\widetilde{\nabla}_{a}v^{d}=\frac{1}{2}\widetilde{g}^{cd}
\{\nabla_{a}h_{cb}+\nabla_{b}h_{ca}-\nabla_{c}h_{ba}\}v^{a}v^{b},
\quad\tilde{ \nabla}_{b}  \tilde{g}_{ac}=0
\end{equation}
where $v^{a}$ is the test--particle four--velocity.  The
(\ref{amiciaddio}) can be written also as
$$
X^{dc}(h_{cb,a}+h_{ca,b}-h_{ba,c})v^{a}v^{b}+X^{dc}(g_{cb,a}+g_{ca,b}-g_{ba,c})v^{a}v^{b}+
$$
\begin{equation}\label{perpieta}
+g^{dc}(h_{cb,a}+h_{ca,b}-h_{ba,c})v^{a}v^{b}=0.
\end{equation}
In particular the geodetic motion in the spacetime endowed with
(\ref{lassassinoiltrucido})  is described by the following equation
\begin{eqnarray}
\nonumber v^{a}\widetilde{C}_{\phantom\ ac}^{b}v^{c} &=&
2(\ln\Omega)_{,v}v^{b}-(g_{ac}v^{a}v^{c})\left[g^{bd}\nabla_{d}\ln\Omega+X^{bd}\nabla_{d}\ln\Omega\right]+
2v^{a}\left(\nabla_{a}\ln\Omega\right)g_{dc}X^{db}v^{c}+
\\ \nonumber +&&\frac{1}{2}v^{a}
\left[\left(\nabla_{a}h_{cd}\right)+\right.\left(\nabla_{c}h_{ad}\right)-
\left.\left(\nabla_{d}h_{ac}\right)\right]v^{c}\left(g^{bd}\Omega^{-2}+X^{bd}\right).
\end{eqnarray}
where $\widetilde{C}_{\phantom\ ac}$ are the tensors defined as$
\widetilde{\nabla}_{a}\omega_{b}=\nabla_{a}\omega_{b}-\widetilde{C}^{c}_{\phantom\
ab}\omega_{c}, \,
\widetilde{\nabla}_{a}t^{b}=\nabla_{a}t^{b}+\widetilde{C}^{b}_{\phantom\
ac}t^{c}$.
\subsection{Lensing}
The light propagation in the $(M,\tilde{g})$ can be described by the
following lensing equation
\begin{equation}\label{25.0}
\frac{dp^{\alpha}}{d\lambda}+\widetilde{\Gamma}^{\alpha}_{\phantom\
\beta\gamma}p^{\beta}p^{\gamma}=0,\quad \mbox{or}\quad\Delta
p^{y}\equiv-\int_{-\infty}^{+\infty}\Gamma^{y}_{\phantom\ \beta
\gamma}p^{\beta}p^{\gamma}\ d\lambda,
\end{equation}
where $\widetilde{\Gamma}^{\alpha}_{\phantom\ \beta\gamma}$ are the
connection coefficients compatible with a $\tilde{g}$ and
$p^{\alpha}\equiv dx^{\alpha}/d\lambda$ is the four--vector wave,
$\lambda$ is a parameter on the geodetic. We assume the trajectory
is on $\texttt{x}\texttt{y}$ plane then, if the motion is initially
parallel to the  $\texttt{x}$ axe, the deflection angle $\Phi$ can
be expressed in terms  of $p^{y}$ as$
\frac{p^{y}}{p^{x}}=\tan\Phi\simeq\Phi$  for $\Phi\simeq0$ and
$p^{x}$ is considered constant {\small{(Cf.~\cite{Wald}})}. We
consider the source very far from the observer and from the lens,
consequently the four--momentum  $p^{y}$ is given by  the following
integral along the trajectory.
$$
\delta
p^{y}=-\frac{\eta_{\mbox{\tiny{QP}}}}{2}\int_{-\infty}^{+\infty}\tilde{g}^{y\delta}
[\partial_{\beta}(\tilde{\omega}^{\mbox{\tiny{Q}}}_{\gamma}\tilde{\omega}^{\mbox{\tiny{P}}}_{\delta})+
\partial_{\gamma} (\tilde{\omega}^{\mbox{\tiny{Q}}}_{\beta}\tilde{\omega}^{\mbox{\tiny{P}}}_{\delta})
-\partial_{\delta}(\tilde{\omega}^{\mbox{\tiny{Q}}}_{\beta}\tilde{\omega}^{\mbox{\tiny{P}}}_{\gamma})]
p^{\beta}p^{\gamma}d\lambda \quad\mbox{or  also,}
$$
\begin{eqnarray}
\nonumber\delta p^{y}&=&\Delta
p^{y}-\frac{1}{2}\int_{-\infty}^{+\infty}[X^{y\delta}(g_{\delta
\beta,\gamma}+g_{\delta \gamma,\beta}-g_{\beta\gamma,\delta}) +
g^{y\delta}(h_{\delta \beta,\gamma}+h_{\delta \gamma,\beta}-h_{\beta
\gamma,\delta})] p^{\beta}p^{\gamma}\ d\lambda+\\
&&\nonumber -\frac{1}{2}\int_{-\infty}^{+\infty}X^{y
\delta}(h_{\delta \beta,\gamma}+h_{\delta
\gamma,\beta}-h_{\beta\gamma,\delta}) p^{\beta}p^{\gamma}\ d\lambda\
\ .
\end{eqnarray}
\subsection{Schwarzschild lensing}
A stationary distribution of matter constituted by a spherical
symmetric body (Schwarzschild lens) {\small{\cite{Lensing}}} is
deformed as follows in the cartesian coordinate
\begin{equation}\label{3.7bis}
\widetilde{ds}^{2}=A\left(1+\frac{2\Phi}{c^{2}}\right)c^{2}dt^{2}-A\left(1-\frac{2\Phi}{c^{2}}\right)\delta_{ij}dx^{i}dx^{j}\,
\end{equation}
In the approximation of weak field, holding if the considered
distances are very much larger then the Schwarzschild radius  of the
field, {\small{$\Phi=-GM/|x|$}}is the gravitational potential
associated to the matter distribution,
$|x|^{2}=\delta_{ij}x^{i}x^{j}$. We obtain the following lensing
equation \cite{Lensing} $
\frac{d\textbf{e}}{dl_{eucl}}=-\frac{2}{c^2}\nabla_{\bot}\Phi $
where
$\nabla_{\bot}\Phi\equiv[\nabla\Phi-\textbf{e}(\textbf{e}.\nabla\Phi)]$,
 $ e^{k} = \frac{dx^{k}}{d \lambda}$ is the spatial vector with
$dl_{eucl}^{2}\equiv\delta_{ij}dx^{i}dx^{j}$. We considered the
following two cases: for $A=A(t)$ the lensing equation became  $
\frac{d\textbf{e}}{dl_{eucl}}=-\frac{2}{c^2}\nabla_{\bot}\Phi-\left(\partial_{0}
\ln A\right)\textbf{e}$; for $A=A(r)$ the equation became $
\frac{d\textbf{e}}{dl_{eucl}}=-\nabla_{\bot}\hat{\Phi} $ where
$\hat{\Phi}\equiv\frac{2}{c^2}\Phi+\ln A $. It is particularly
interesting to note that the last transformation, that is nothing
more than a diffeomorphism, reduces itself, in the lensing effect to
a redefinition of the gravitational potential.
\subsection{Geodetic deviation}
 A family   $\gamma_{s}(t)$  at one parameter  $s$ of
geodetics of $(M,\widetilde{g})$ is parameterized by the affine
parameter  $t$ in such a way that a point $p\in\gamma_{s}(t)$ is
well defined by the map $(t,s)\rightarrow\gamma_{s}(t)$.The
separation among the two curves changes respect to the parameter $t$
with a velocity $v^{a}\equiv T^{b}\widetilde{\nabla}_{b}S^{a}$ given
by the geodesic equation:
\begin{equation}\label{25.10}
a^{a}\equiv
T^{c}\widetilde{\nabla}_{c}v^{a}=-\widetilde{R}_{cbd}^{\phantom{cbd}
a}S^{b}T^{c}T^{d}=a_{0}^{a}-\overline{R}_{cbd}^{\phantom\ \ \
a}S^{b}T^{c}T^{d}, \quad a_{0}^{a}\equiv -R_{cbd}^{\phantom\ \ \
a}S^{b}T^{c}T^{d}.
\end{equation}
where  $T^{a}=\left(\frac{\partial}{\partial t}\right)^{a}$ is
four--vector tangent to the family of geodetics  and the
four--vector $S^{b}=\left(\frac{\partial}{\partial s}\right)^{b}$
gives the infinitesimal displacement among two infinitesimal close
geodetics, $R_{cbd}^{\phantom\ \ \  a}$ is the curvature  tensor. We
finally find
\begin{equation}\label{fermaferma}
a^{a}=-2 (\partial_{[b}\widetilde{C}^{a}_{\phantom\
|d|c]}+\widetilde{C}^{f}_{\phantom\
[c|d|}\widetilde{\Gamma}^{a}_{\phantom\ b]f}+\Gamma^{f}_{\phantom\
[c|d|}\widetilde{C}^{a}_{\phantom\  b]f})S^{b}T^{c}T^{d}.
\end{equation}
\subsection{Gravitational redshift} Consider two observers $O_{1}$
and  $O_{2}$  at rest in the spacetime $(M,\widetilde{g})$, where
$\widetilde{g}_{ab}=\mathcal{G}_{\mbox{\tiny{AB}}}\omega_{a}^{\mbox{\tiny{A}}}\omega_{b}^{\mbox{\tiny{B}}}$
is a static metric $(-,+,+,+)$. $O_{1}$    emits at  $P_{1}$ a light
signal received by  $O_{2}$ at $P_{2}$. In the approximation of the
geometric optic the light rays propagate in the spacetime
$(M,\widetilde{g})$ along null geodetic. The frequency  $\nu$, the
wave four--vector $p^{a}$and the four--velocity $v^{a}$ are linked
by the relation: $ \nu=-p_{a}v^{a}. $ From  this we find the
emission frequency $\nu_{1}$ and the frequency of absorbing
$\nu_{2}$ of the photons traveling between the two observers $
\nu_{i}=-(p_{a}v^{a}_{i})|_{P_{i}}$ where   $v^{a}_{1}$  and
$v^{a}_{2}$  are the four--velocity of the observers  $O_{1}$  and
$O_{2}$  respectively. Because of the  vectors  $v^{a}_{1}$  and
$v^{a}_{2}$ are tangent to the Killing time--vector field $\xi^{a}$
$(\xi^{a}\xi_{a}=\widetilde{g}_{00})$,we find : $
v_{i}^{a}=[\xi^{a}/(-\xi^{b}\xi_{b})^{½}]|_{P_{i}}$ Then using the
property that the internal product between the vector field
$\xi^{a}$ and the tangent field $p^{a}$ is constant along the
geodetic we  have in terms of the wave lengths   $\lambda_{1}$  and
$\lambda_{2}$:
\begin{equation}\label{echiusolaperche}
z\equiv\frac{\lambda_{2}-\lambda_{1}}{\lambda_{1}};\quad
z=\frac{\nu_{1}}{\nu_{2}}-1=\frac{\sqrt{-\mathcal{G}_{\mbox{\tiny{AB}}}\omega^{\mbox{\tiny{A}}}_{0}\omega^{\mbox{\tiny{B}}}_{0}}|_{P_{2}}}
{\sqrt{-\mathcal{G}_{\mbox{\tiny{AB}}}\omega^{\mbox{\tiny{A}}}_{0}\omega^{\mbox{\tiny{B}}}_{0}}|_{P_{1}}}
-1.
\end{equation}
\section{Conclusion}
In this work we deformed spacetime metrics using  matrices of scalar
fields. Such deformations generalize conformal transformations. In
particular we considered deformations as a generalization of the
Gauss theorem to every couple n-dimensional metrics. We focused on
the main features of deformed manifold showing, in particular, in
what sense deformations differ by diffeomorphism. The equation of
gravitational waves in the case of ``small" deformation is obtained.
Finally, we recover the equations of motion in deformed spacetime
studying, in particular, the lensing effects and gravitational
redshift induced by deformations. Details on the gravitational lens
produced by  deformations of the Schwarzschild metric are given, in
particular a conformal deformation of the metric induces a
deformation of the gravitational potential. The equation of geodetic
deviation is also obtained. These considerations motivated  the idea
that deformations can be  used to adapt  solutions of Einstein's
equations  to the cosmological observations and more generally to
the experimental data.


\begin{thebibliography}{9}
\bibitem{Storny}
S.Capozziello, C. Stornaiolo {\it Int.J.Geom.Meth.Mod.Phys,}
(05),185-195 (2008).
 %%
%\bibitem{4--dim} B. Coll
%\emph{A Universal Law of Gravitational Deformation for General
%Re\-la\-ti\-vi\-ty} in Proceedings oh the Spanish Relativity meeting
%in honour of 65th birthday of Luis Bell ``Gravitation and Relativity
%in General'' ed by Martin, J. Ruiz, E. Atrio, F. Molina, A. World
%Scientific.(1999)
%%
\bibitem{3--dim} B. Coll, J. Llosa and D. Soler
\emph{Gen. Rel. Grav.}(34), 2 (2002).
%
\bibitem{Llosa e Soler}
J. Llosa and D. Soler, {\it Class. Quant. Grav.}(22), 893 (2005).
%
\bibitem{Lensing} A. A. Marino, S. Capozziello, R. de Ritis and
P. Scudellaro {\it Lezioni di Lensing Gravitazionale} (Bibliopolis,
2000).
%Vedi come si scrive che è una tesi si Lauerea
\bibitem{Tesi}
D. Pugliese,{\it Deformazioni di metriche spaziotemporali} Thesis
(unpublished) .   Supervisors: S. Capozziello and C. Stornaiolo.
Universit\`{a} degli studi di Napoli  ``Federico II''. Dipartimento
di Scienze Fisiche, Biblioteca \emph{Roberto Stroffolini} (2007).
\bibitem{Wald}
R. Wald, {\it General Relativity} (The University of Chicago Press,
1984).
%Article:
%\bibitem{label}
%F. Author and S. Author, {\it Jour. Name} {\bf 55}(22), 25 (2008).
%
%%Book:
%\bibitem{labe1}
%F. Mandl and G. Shaw, {\it Quantum Field Theory} (J. Wiley \& Sons, Chichester (UK), 1998).
%
\end{thebibliography}
\end{document}